\documentclass[fleqn,12pt,twoside]{article}
\usepackage{espcrc1}

\input BoxedEPS
\SetRokickiEPSFSpecial  
\HideDisplacementBoxes

\def\rd#1{\mathop{{\rm d}#1}}
\def\braket#1#2#3{\bigl\langle #1 \bigl| #2 \bigr| #3\bigr\rangle}

\newcommand{\AmS}{{\protect\the\textfont2
  A\kern-.1667em\lower.5ex\hbox{M}\kern-.125emS}}

\title{Understanding Hadron Structure Using Lattice QCD\thanks{Work supported in
part by the U.S. Department of Energy (DOE) under cooperative research agreement
\#DE-FC02-94ER40818. \quad MIT-CTP-3162}}

\author{ J. W. Negele\address
{Center for Theoretical Physics, 
Massachusetts Institute of Technology, \\ 
        77 Massachusetts Avenue, Cambridge, Massachusetts 02139, USA}}

\begin{document}

\maketitle

\begin{abstract}
Numerical evaluation of the path integral for QCD on a
discrete space-time lattice has been used to calculate
ground state matrix elements specifying moments of quark
density and spin distributions. This talk will explain how
these matrix elements have been calculated in full QCD using
dynamical quarks,  show how physical extrapolation to the
chiral limit including  the physics of the pion cloud
resolves previous apparent conflicts with experiment, and 
describe the computational resources required for a
definitive comparison with experiment.

\end{abstract}

\section{INTRODUCTION}

The structure of hadrons differs profoundly from that of other familiar many body
systems. Unlike electrons in atoms or nucleons in nuclei, quarks are confined in the
nucleon. Whereas photons can be subsumed into the Coulomb potential in atomic
physics and pions can be subsumed into the nucleon-nucleon potential in nuclear
physics, gluons are essential degrees of freedom in light hadrons that carry half the
momentum and have important nonperturbative topological excitations.  Since
nonperturbative QCD is presently intractable analytically, the goal of this work is to
use lattice field theory to solve QCD with controlled errors to provide a quantitative
understanding of the rich quark and gluon structure  of the nucleon and to obtain
insight into how QCD actually works to produce this structure.

Decades of high energy scattering experiments have now provided 
detailed experimental knowledge of the light cone
distributions of quarks and gluons in the nucleon, so this work addresses the use of
lattice QCD to understand the observed distribution of the quark density and helicity.
Using the
operator product expansion, it is possible to calculate
moments of quark distributions, and I will discuss here the first
calculations in full QCD\cite{dolgov-thesis,Dolgov:2001ca}. A major puzzle in the field
has been the fact that quenched calculations of these moments, which ignore
quark-antiquark excitations of the Dirac sea, disagree with experiment at the
20-50\% level. I will show that contrary to some conjectures, at the quark masses
accessible in practical calculations, including quark loops does not alter the results
significantly. Rather,  I will  argue that the physical origin of the discrepancy with
experiment has been incorrect extrapolation to the physical quark mass, and will show
how extrapolation incorporating the leading non-analytic behavior required by chiral
symmetry produces consistent results for the moments of quark distributions. In
addition, we have  also  compared full QCD results with  configurations that have been
cooled to remove all the gluon contributions except for those of instantons and shown
that the qualitative behavior of the moments is reproduced by the instanton content
of the gluon configurations.  

\section{MOMENTS OF QUARK DISTRIBUTIONS IN THE PROTON}

\begin{figure}[tp]
\vspace*{-2mm}
$$
\BoxedEPSF{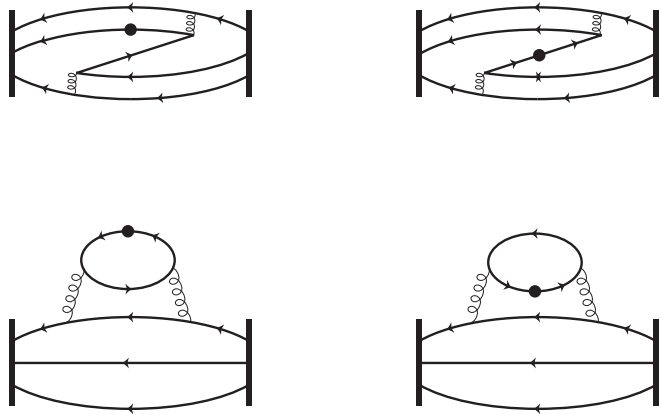 scaled 900}
$$
\vspace*{-.6cm }
\caption{Connected (upper row) and disconnected (lower row) diagrams contributing
to hadron matrix elements. The left column shows typical contributions of quarks and
the right column shows contributions of antiquarks}
\label{HadrMatrElem}  
\vspace*{-.5cm} 
\end{figure}

By the operator product expansion, moments of the following linear combinations of
quark and antiquark distributions in the proton
\begin{eqnarray}
\langle x^n\rangle_q &=&
\int_0^1 \rd x x^n \bigl(q (x) + (-1)^{n+1}\bar q(x)\bigr)  \label{mom1}\\
\langle x^n\rangle_{\Delta q} &=&
\int_0^1 \rd x x^n \bigl(\Delta q (x) + (-1)^{n}\Delta\bar q(x)\bigr)  \nonumber\\
\langle x^n\rangle_{\delta q} &=& 
\int_0^1 \rd x x^n \bigl(\delta q (x) + (-1)^{n+1}\delta\bar q(x)\bigr) \nonumber
 \end{eqnarray}
are related to the following matrix elements of twist-2 operators
\begin{eqnarray}
\braket {PS}{\bar\psi \gamma^{\{\mu_1} i  D^{\mu_2} \cdots i
D^{\mu_n\}}
\psi}{PS} &=& 2 \langle x^{n-1}\rangle_q\, P^{\{\mu_1}\cdots P^{\mu_n\}} 
\label{mom2}\\
\braket {PS}{\bar\psi \gamma^{\{\mu_1}\gamma_5 i  D^{\mu_2} \cdots i
D^{\mu_n\}}
\psi}{PS} &=& 2 \langle x^{n-1}\rangle_{\Delta q}\,
MS^{\{\mu_1}P^{\mu_2}\cdots P^{\mu_n\}} \nonumber\\
\braket{PS}{\bar\psi \sigma^{[\alpha\{\mu_1]}\gamma_5 i 
D^{\mu_2}
\cdots i D^{\mu_n\}}
\psi}{PS} &=& 2 \langle x^{n-1}\rangle_{\delta q}\,
MS^{[\alpha}P^{\{\mu_1]}P^{\mu_2}\cdots P^{\mu_n\}} \nonumber.
\end{eqnarray}
Here,   $q = q_\uparrow + q_\downarrow ,  \Delta q =
q_\uparrow - q_\downarrow$,  $\delta q = q_\top + q_\bot $, $x$ denotes the
momentum fraction carried by the quark, and  $\{\,\} $ and $[\,] $ denote
symmetrization and  antisymmetrization respectively. We note that odd moments 
$\langle x^n\rangle_q$ are obtained from deep inelastic electron or muon scattering
structure functions
$F_1$ or
$F_2$,  even moments of $\langle x^n\rangle_{\Delta q} $ are determined from
$g_1$, and these moments are proportional to the quantities $v_{n+1}$ and $a_n$
defined in ref.\cite{qcdsf}. In addition, $g_2$ also determines the quantity $d_n$
\begin{equation}
\langle PS |  \bar{\psi}
\gamma^{[\sigma}\gamma_5iD^{\{\mu_1]} \cdots
iD^{\mu_n\}}\psi | PS\rangle = \frac{1}{n} {d_n
S^{[\sigma}P^{\{\mu_1]}\cdots P^{\mu_n\}}} 
\end{equation}
Even moments  $\langle
x^n\rangle_q$ are obtained from deep inelastic neutrino scattering, and in addition, a
variety of other processes have contributed to what is now a detailed empirical
knowledge of the quark and antiquark distributions in the nucleon. We will
subsequently compare our  results with moments calculated from the CTEQ, MRS, and
GRV global fits to the world supply of data.

\section{CALCULATION OF MATRIX ELEMENTS}

Proton matrix elements of the operators in Eq.\ref{mom2} are calculated by
evaluating the connected and disconnected diagrams shown in Fig.
\ref{HadrMatrElem}. Note that both the connected and disconnected diagrams each
receive contributions from quarks and antiquarks. Depending on the moment, by
Eq.\ref{mom1}, the sum of the diagrams yields either the sum or difference of the
moments of the quark and antiquark distributions. In the past, there has been some
confusion on this point: because the connected diagrams are called the valence quark
distribution and experimentalists define the valence quark distribution as the
difference between the quark and antiquark distributions, the connected
contributions have erroneously been compared only with the difference between the
empirical quark and antiquark distributions. Because it is technically much more
difficult to evaluate the disconnected diagrams, our present calculations only include
connected diagrams. Fortunately, the disconnected diagrams are flavor independent,
so they cancel out of the difference between up and down quark distributions. Hence,
in Table
\ref{tab-summary},  we compare lattice calculation of the difference
between connected diagrams for up and down quarks with the corresponding
difference  of moments of experimental data for the sum or difference of quark and
antiquark distributions.
%
%

\begin{figure}[tp]
\vspace*{-2mm}
$$
\BoxedEPSF{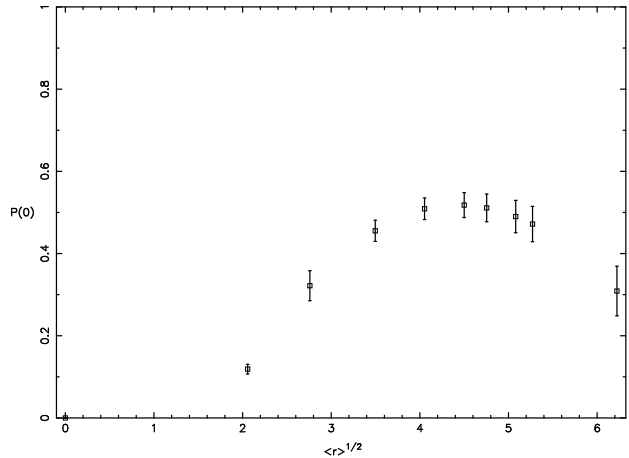 scaled 470}
$$
\vspace*{-3pc}
\caption{Overlap between smeared source and proton ground state as a function of
the source RMS radius. The overlap for zero smearing is approximately $10^{-4}$}
\label{overlap}  
\vspace*{-.5cm} 
\end{figure}

\begin{figure}[tp]
\vspace*{-2mm}
$$
\BoxedEPSF{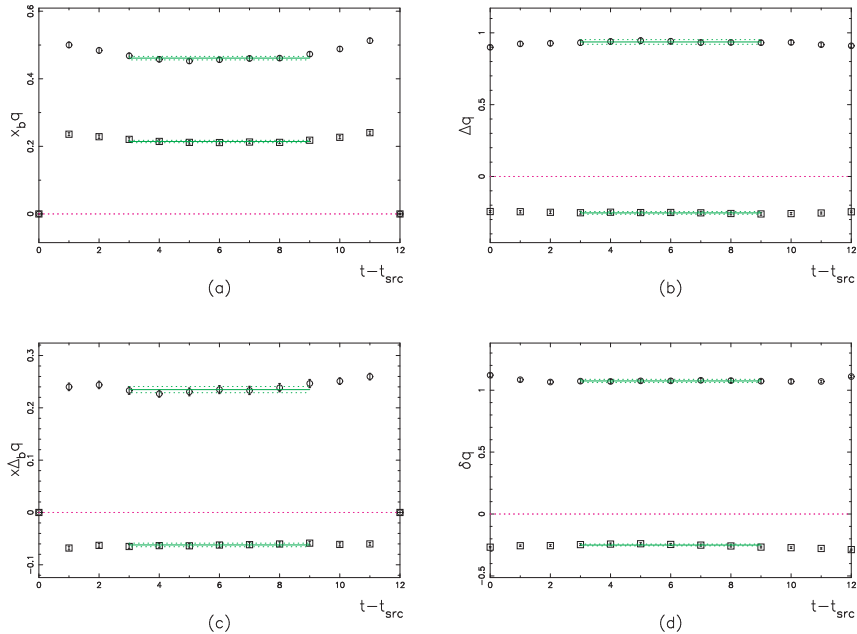 scaled 450}
$$
\vspace*{-3pc}
\caption{Plateaus in measurements of operators in a zero momentum ground state as
a function of the euclidean time separation from the source.}
\label{NewFigs1}  
\vspace*{-.75cm} 
\end{figure}

On the lattice, connected diagrams are evaluated by calculating a three point function
in which a source creates a state with the quantum numbers of the proton, 
the operator acts on this state, and a sink finally annihilates the state. 
Because evolution in imaginary time filters out the ground state,  when the
operator is sufficiently far from both the source and sink, it acts in the
ground state and produces the desired ground state matrix element. As the time at
which it acts approaches either the source or sink, it sees excited state contaminants,
yielding a central plateau corresponding to the physical matrix element and
exponential contaminants at the edges. Obviously, it is beneficial to optimize the
overlap of the source with the ground state to maximize the plateau region and
minimize the effect of the excited state contaminants at the edges.

In this work, connected diagrams were calculated using sequential
propagators generated by the upper two components of the
nucleon source $J^{\alpha} =
u_a^{\alpha}u_b^{\beta}(C\gamma_5)_{\beta,\beta'}
d_c^{\beta'}\epsilon^{abc}$. The overlap with the physical
proton ground state was optimized using Wuppertal
smearing \cite{sources} to maximize the  overlap  $P(0) = {|\langle J|0\rangle |^2}$.
Figure~\ref{overlap}
shows that varying the smearing reduced $P$ by over 4
orders of magnitude, yielding an overlap with the physical
ground state of approximately 50\%. Dirichlet boundary
conditions were used for quarks in the t-direction. 

The resulting plateaus for four operators that could be measured in a proton with
zero three-momentum are shown in Fig.~\ref{NewFigs1}. Here one observes both a
statistically well determined central plateau region and the effectiveness with which
the excited state contaminants have been reduced by the optimized source.
\vspace*{-\medskipamount}

\section{OPERATORS AND
PERTURBATIVE RENORMALIZATION}

The continuum operators defined above are approximated on a
discrete cartesian lattice using representations of the
hypercubic group that have been chosen to eliminate operator mixing as much as
possible and to  minimize the number of non-zero
components of the
nucleon momentum.  The
operators we have used are shown in Table~1, 
where we
have indicated whether the spatial momentum components
are non-zero and whether mixing occurs. Note, no$^*$
indicates a case in which mixing could exist in general but
vanishes perturbatively for Wilson or overlap fermions and
no$^{**}$ indicates perturbative mixing with {\it lower}
dimension operators for Wilson fermions but no mixing for
overlap fermions.  Because the statistical errors are much larger for sources
projected to non-zero momentum, the moments corresponding to operators requiring
non-zero momentum are presently not well determined.

To convert from  lattice regularization at the scale of the inverse lattice spacing $1/a$
to the continuum $\overline{MS}$ scheme at momentum scale Q, we use the one-loop
perturbation theory result
\begin{displaymath}
O^{\overline{MS}}_i(Q^2)=\sum_j\left(\delta_{ij}+\frac{g_0^2}{16\pi^2}\,
\frac{N_c^2-1}{2N_c}
\left(\gamma^{\overline{MS}}_{ij}\log(Q^2a^2)-(B^{LATT}_{ij}
-B^{\overline{MS}}_{ij})\right)\right)\cdot O^{LATT}_j(a^2) .
\label{app-renorm1}
\end{displaymath}
The anomalous dimensions $\gamma_{ij}$ and the finite constants $B_{ij}$
we have
calculated and used in this work are tabulated in
Table~2 \cite{pert-renorm}. 
\bigskip

\leavevmode\kern-3pc
\begin{tabular}{cc}
\stepcounter{table}
\noindent
\begin{minipage}[t]{0.5\hsize}
Table 1\quad Lattice Operators\\ 
\begin{small}
\begin{tabular}{llrl}
\hline
& H(4) mix&$\vec{p}$&lattice operator\\
\hline
$xq_c^{(a)}$ & {\bf 6}$_3^+$ \, no & $\neq 0$  & $\bar{q} \gamma_{\{1} 
{\stackrel{\,\leftrightarrow}{D}}_{4\}} {q}$  \\
$xq_c^{(b)}$ & {\bf 3}$_1^+$ \, no & 0& $\bar{q} \gamma_{4} 
{\stackrel{\,\leftrightarrow}{D}}_{4} {q}$ \\
& & \multicolumn{2}{c}{ $~~~- \frac{1}{3} \sum_{i=1}^3
 \bar{q} \gamma_{i} {\stackrel{\,\leftrightarrow}{D}}_{i} {q}$}  \\
$x^2q_c$     & {\bf 8}$_1^-$ \, yes & $\neq 0$ & $\bar{q} \gamma_{\{1} 
{\stackrel{\,\leftrightarrow}{D}}_{1} 
{\stackrel{\,\leftrightarrow}{D}}_{4\}} {q}$ \\
& & \multicolumn{2}{c}{ $~~~- \frac{1}{2} \sum_{i=2}^3
\bar{q}\gamma_{\{i} {\stackrel{\,\leftrightarrow}{D}}_{i} 
{\stackrel{\,\leftrightarrow}{D}}_{4\}}{q}$} \\
$x^3q_c$     & {\bf 2}$_1^+$ \, no$^*$ & $\neq 0$ & $\bar{q} \gamma_{\{1} 
{\stackrel{\,\leftrightarrow}{D}}_{1} 
{\stackrel{\,\leftrightarrow}{D}}_{4} 
{\stackrel{\,\leftrightarrow}{D}}_{4\}} {q}$ \\
& & \multicolumn{2}{c}{ $~~~+ \bar{q} \gamma_{\{2} {\stackrel{\,\leftrightarrow}{D}}_{2} 
{\stackrel{\,\leftrightarrow}{D}}_{3} 
{\stackrel{\,\leftrightarrow}{D}}_{3\}} {q}$} \\
& & \multicolumn{2}{c}{ $~~~- (\,3\,\leftrightarrow\,4\,)$} \\
\hline
$\Delta q_c$     & {\bf 4}$_4^+$ \, no & 0  & $\bar{q} 
\gamma^{5} \gamma_{3} {q}$  \\
$x\Delta q_c^{(a)}$ & {\bf 6}$_3^-$  \, no & $\neq 0$ & $\bar{q} 
\gamma^5 \gamma_{\{1} {\stackrel{\,\leftrightarrow}{D}}_{3\}} {q}$ \\
$x\Delta q_c^{(b)}$ & {\bf 6}$_3^-$  \, no & 0 & $\bar{q} \gamma^5 
\gamma_{\{3} {\stackrel{\,\leftrightarrow}{D}}_{4\}} {q}$ \\
$x^2\Delta q_c$     & {\bf 4}$_2^+$  \, no & $\neq 0$ & $\bar{q} 
\gamma^5 \gamma_{\{1} {\stackrel{\,\leftrightarrow}{D}}_{3} 
{\stackrel{\,\leftrightarrow}{D}}_{4\}} {q}$ \\
\hline
$\delta q_c$     & {\bf 6}$_1^+$  \, no & 0  & $\bar{q} \gamma^{5} \sigma_{34} 
{q}$  \\
$x\delta q_c$    & {\bf 8}$_1^-$ \, no & $\neq 0$   & $\bar{q} \gamma^{5} 
\sigma_{3\{4} {\stackrel{\,\leftrightarrow}{D}}_{1\}}{q}$  \\
$d_1$     & {\bf 6}$_1^+$  \, no$^{**}$ & 0  & $\bar{q} \gamma^5 \gamma_{[3} 
{\stackrel{\,\leftrightarrow}{D}}_{4]}{q}$ \\
$d_2$     & {\bf 8}$_1^-$  \, no$^{**}$ & $\neq 0$ & $\bar{q} \gamma^5 
\gamma_{[1} {\stackrel{\,\leftrightarrow}{D}}_{\{3]} 
{\stackrel{\,\leftrightarrow}{D}}_{4\}} {q}$ \\
\hline
\end{tabular}
\end{small}
\end{minipage}
&
\stepcounter{table}
\begin{minipage}[t]{0.5\hsize}
Table 2\quad Perturbative renormalization\\[-.5ex]
\hphantom{Table 2\quad}constants\\[1ex]
\setlength{\tabcolsep}{5pt}
\begin{small}
\begin{tabular}{lcrcll}
\hline
& $\gamma$ & $B^{LATT}$ & $B^{{\overline{MS}}}$ & \multicolumn{2}{c}{$Z$} \\
& & & & ${}_{\beta=6.0}$&${}_{\beta=5.6}$ \\
\hline
$xq^{(a)}$         &     $\frac{8}{3}$  &   $-3.16486$  &      $-\frac{40}{9}$
   & $0.989$& $0.988$ \\
$xq^{(b)}$         &     $\frac{8}{3}$  &   $-1.88259$  &      $-\frac{40}{9}$
   & $0.978$& $0.977$ \\
$x^2q$             &    $\frac{25}{6}$  &  $-19.57184$  &      $-\frac{67}{9}$
   & $1.102$& $1.110$ \\
$x^3q$             &  $\frac{157}{30}$  &  $-35.35192$  &  $-\frac{2216}{225}$
   & $1.215$& $1.231$ \\
$\Delta q$         &               $0$  &   $15.79628$  &                  $0$
   & $0.867$& $0.857$ \\
$x\Delta q^{(a)}$  &     $\frac{8}{3}$  &   $-4.09933$  &      $-\frac{40}{9}$
   & $0.997$& $0.997$ \\
$x\Delta q^{(b)}$  &     $\frac{8}{3}$  &   $-4.09933$  &      $-\frac{40}{9}$
   & $0.997$& $0.997$ \\
$x^2\Delta q$      &    $\frac{25}{6}$  &  $-19.56159$  &      $-\frac{67}{9}$
   & $1.102$& $1.110$ \\
$\delta q$         &               $1$  &   $16.01808$  &                 $-1$
   & $0.856$& $0.846$ \\
$x\delta q$        &               $3$  &   $-4.47754$  &                 $-5$
   & $0.996$& $0.995$ \\
$d_1$              &               $0$  &  $0.36500$  &                  $0$
   & $0.997$& $0.997$ \\
$d_2$              &     $\frac{7}{6}$  &  $-15.67745$  &     $-\frac{35}{18}$
   & $1.116$& $1.124$ \\
\hline
\end{tabular}
\end{small}
\end{minipage}
\end{tabular}

\begin{figure}[tp]
\vspace*{-2mm}
$$
\BoxedEPSF{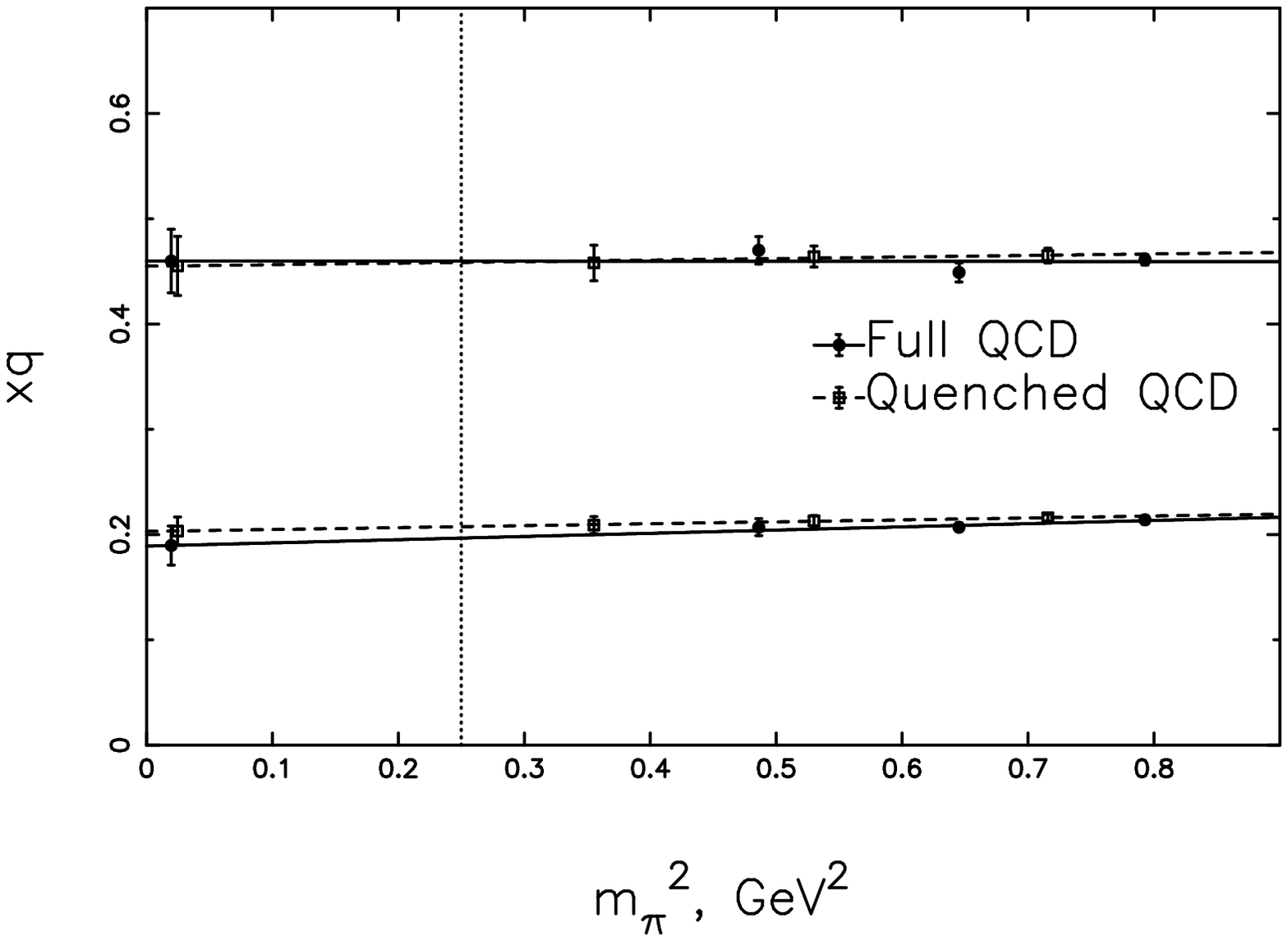 scaled 500}
\quad
\BoxedEPSF{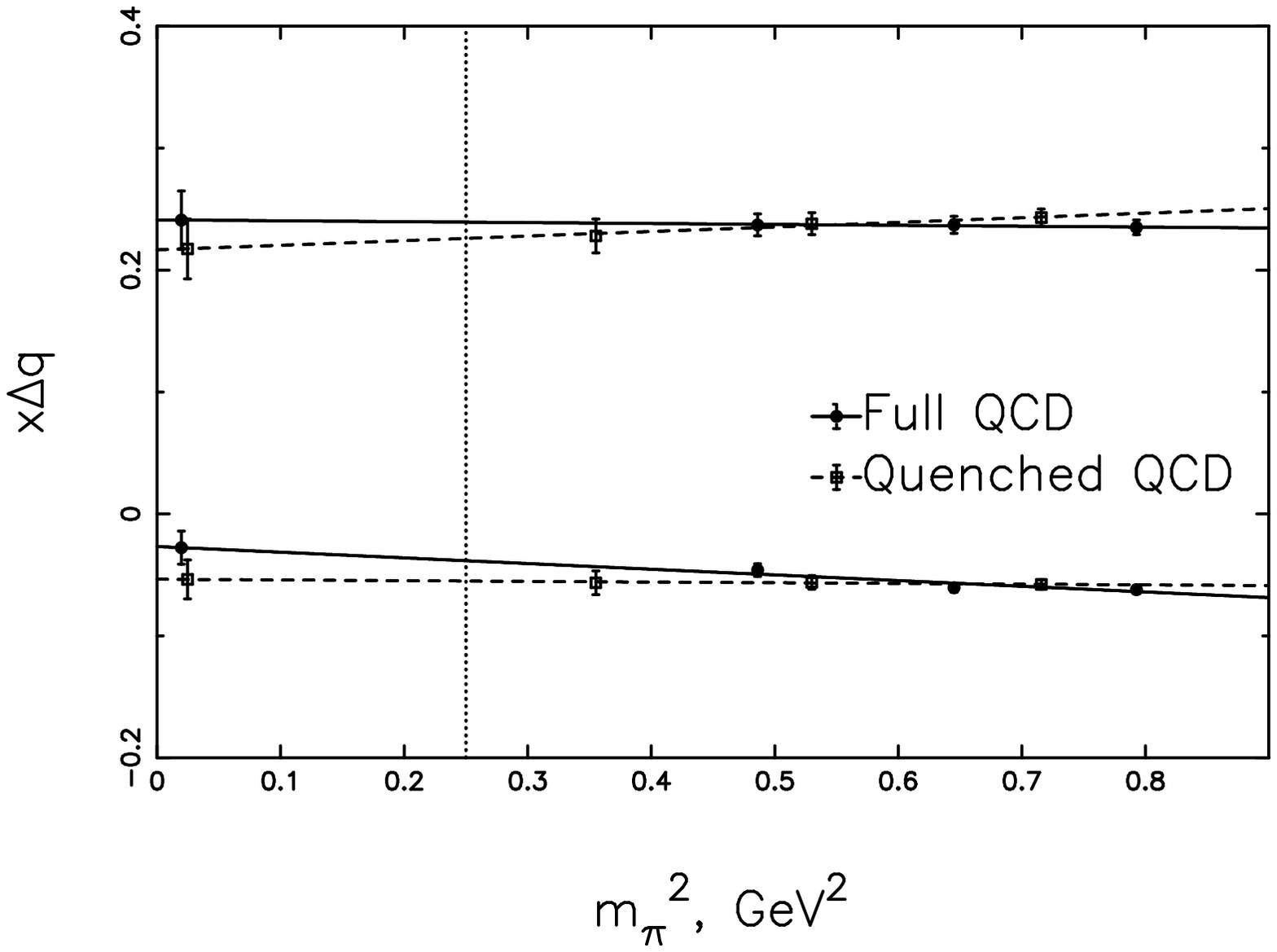 scaled 500}
$$
\vspace*{-4pc}
\caption{Comparison of linear chiral extrapolations of full and quenched
calculations of $\langle x q \rangle$ and  $\langle x \Delta q
\rangle$  showing agreement within statistical errors  }
 \label{fig-full}
\vspace*{-.5cm} 
\end{figure}

\begin{figure}[tp]
\vspace*{-4mm}
$$
\BoxedEPSF{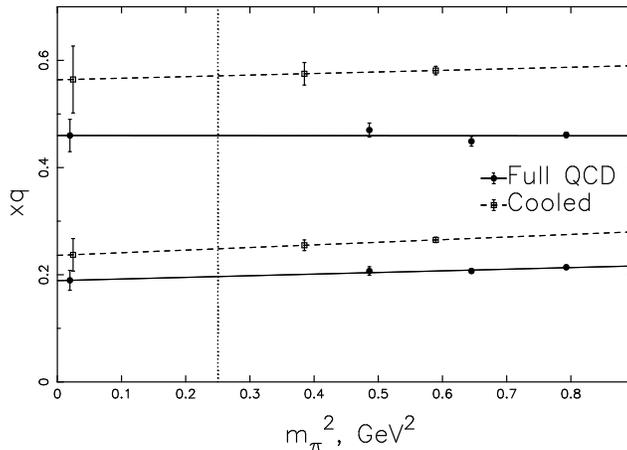 scaled 450}
$$
\vspace*{-3.5pc}
\caption{Comparison of linear chiral extrapolations of full and cooled
calculations of $\langle x q \rangle$ showing the extent to
which instantons reproduce the full result.}
\label{fig-full-cooled}
\vspace*{-.5cm} 
\end{figure}

\begin{table}[hbt]
\caption{Comparison of linear extrapolations of our full QCD and quenched results with
other lattice calculations and phenomenology at  $4$ GeV in $\overline{MS}$}
\label{tab-summary}
\begin{center}
\begin{footnotesize}
\begin{tabular}{lrrrrrr}
\hline
Connected & QCDSF & QCDSF & Wuppertal &Quenched &Full QCD 
& \multicolumn{1}{c}{Phenomenology} \\
 M. E.&  & ($a=0$) &  &  & (3 pts) & \multicolumn{1}{c}{($q\pm \bar q$)} \\  
\hline
$ \langle x \rangle_u $      & $0.452(26)$ &   &   & $0.454(29)$ & $0.459(29)$ &  \\
$\langle x \rangle_d $      & $0.189(12)$ &   &   & $0.203(14)$ & $0.190(17)$ &  \\
$\langle x \rangle_{u-d}$    & $0.263(17)$ &   &   & $0.251(18)$ & $0.269(23)$ &
$0.154$ \\ 
$\langle x^2 \rangle_u $    & $0.104(20)$ &   &   & $0.119(61)$ & $0.176(63)$ &  \\
$ \langle x^2 \rangle_d $    & $0.037(10)$ &   &   & $0.029(32)$ & $0.031(30)$ & 
\\
$\langle x^2 \rangle_{u-d} $ & $0.067(22)$& &  & $0.090(68)$ &  $0.145(69) $
& $0.055 $\\
$ \langle x^3 \rangle_u $    & $0.022(11)$ &   &   & $0.037(36)$ & $0.069(39)$ &
 \\
$ \langle x^3 \rangle_d $    & $-0.001(7)$ &   &   & $0.009(18)$ & $-0.010(15)$
 \\
$\langle x^3 \rangle_{u-d} $ & $0.023(13)$& &  & $0.028(49)$ &  $0.078(41) $
& $0.023 $\\
$ \langle 1\rangle_{\Delta u}$ & $0.830(70)$ & $0.889(29)$ & $0.816(20)$
& $0.888(80)$ & $0.860(69)$ &  \\
$\langle 1\rangle_{\Delta d}$ & $-0.244(22)$ & $-0.236(27)$ & $-0.237(9)$
& $-0.241(58)$ & $-0.171(43)$ &  \\
$ \langle 1\rangle_{\Delta u -\Delta d} $ & $1.074(90)$ & $1.14(3)$ & $1.053(27)$  
& $1.129(98)$ & $1.031(81)$ & $1.257$ \\
$\langle x \rangle_{\Delta u} $ & $0.198(8)$ &   &  & $0.215(25)$ & $0.242(22)$ & \\
$\langle x \rangle_{\Delta d}$ & $-0.048(3)$ &   &  & $-0.054(16)$ &
$-0.029(13)$ & \\
$\langle x \rangle_{\Delta u -\Delta d}$ & $0.246(9)$ &      &  & $0.269(29)$ &
$0.271(25)$ & $0.191$ \\
$ \langle x ^2\rangle_{\Delta u} $ & $0.087(14)$ &      &  & $0.027(60)$ & $0.116(42)$
& \\
$ \langle x ^2\rangle_{\Delta d} $ & $-0.025(6)$ &      &  & $-0.003(25)$ &
$0.001(25)$ &\\
$ \langle x ^2\rangle_{\Delta u - \Delta d} $ & $0.112(15)$ &      &  & $0.030(65)$ &
$0.115(49)$ & $0.061$ \\
$ \delta u_c $ & $0.93(3)$ & $0.980(30)$ &   & $1.01(8)$ & $0.963(59)$ &  \\
$ \delta d_c $ & $-0.20(2)$ & $-0.234(17)$ &   & $-0.20(5)$ & $-0.202(36)$ &  \\
$ d_2^u $ & $-0.206(18)$ &     &   & $-0.233(86)$ & $-0.228(81)$ &  \\
$ d_2^d $ & $-0.035(6)$ &     &   & $0.040(31)$ & $0.077(31)$ &  \\
\hline
\end{tabular}
\end{footnotesize}
\end{center}
\vspace*{-3pc}
\end{table}

\section{RESULTS}

\begin{figure}[tp]
\vspace*{-2mm}
$$
\BoxedEPSF{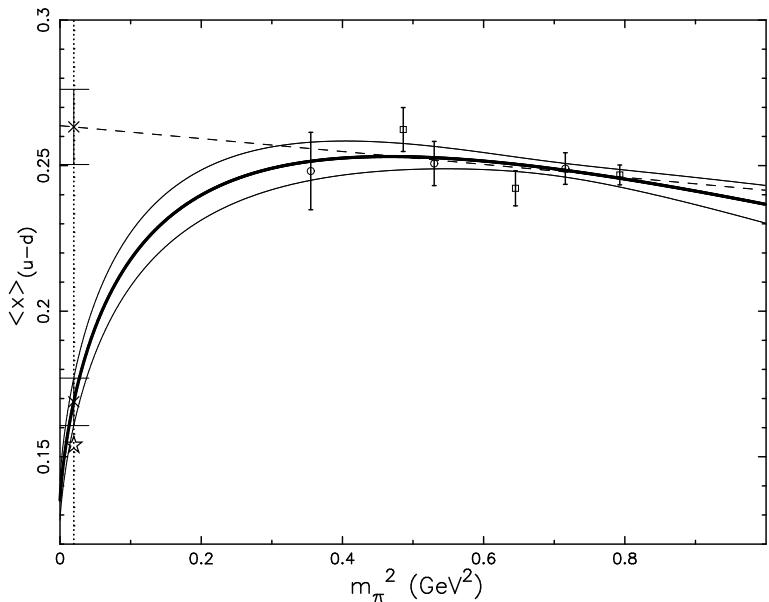 scaled 550}
$$
\vspace*{-3.5pc}
\caption{Chiral extrapolation of $\langle x \rangle_{u-d}$ using Eq. \ref{chiral_ext}}
\label{fit}  
\vspace*{-.9cm} 
\end{figure}

The moments listed in Table 1 were calculated \cite{dolgov-thesis} 
on $16^3 \times 32$ lattices for Wilson 
fermions in full QCD at $\beta = 5.6$ using 200 SESAM configurations
at each of 4 $\kappa 's$ and at $\beta = 5.5$ using 100 SCRI
configurations at 3 $\kappa 's$. 
They were also calculated with two sets of 100 full
QCD configurations cooled with 50 cooling steps and in
quenched  QCD at $\beta = 6.0$ using 200 configurations at each of 3
$\kappa 's$.  Typical linear chiral extrapolations for operators calculated
with nucleon momentum equal to zero are shown in Figure~\ref{fig-full}
for full and quenched
calculations of $\langle x \rangle_q$ and  $\langle x 
\rangle_{\Delta q}$, showing agreement within statistical errors. To avoid
finite volume errors at the 
lightest quark mass, the SESAM \cite{sesam} results were extrapolated using the
three heaviest quark masses. Table~3 shows a major result of our work,
that there is complete
agreement within statistics between full and quenched
results. Statistics with the SCRI configurations \cite{scri} 
are not yet adequate
to present extrapolations in the coupling constant.

Typical 
chiral extrapolations for cooled configurations are compared with the
corresponding uncooled full QCD calculations in
Figure~\ref{fig-full-cooled}. 
This qualitative agreement between cooled and uncooled results occurs at
light quark mass for
all the twist-2 matrix elements we calculated  and demonstrates the
degree to which the instanton content of the configurations 
and their associated zero modes dominate light hadron structure 
\cite{instantons}.

A longstanding puzzle in this field has been the fact that when quenched lattice
results are linearly extrapolated in $m_q$,  results disagree at the 20-50\% level.
Our results show that inclusion of quark-antiquark excitation from the Dirac Sea does
not resolve this discrepancy.  Salient examples from Table \ref{tab-summary}	are
$\langle x \rangle_{u-d} \sim 0.25 - 0.27$  compared with the experimental result
0.15 and $ g_A =
\langle 1\rangle_{\Delta u -\Delta d}  \sim 1.0 - 1.1$  compared with the experimental
result 1.26.

There is strong evidence that the physical origin of these discrepancies is the
inadequate treatment of the pion cloud in the nucleon that has been necessary
because of limited computational resources. By necessity, present calculations are
restricted to quark masses that are so heavy that the pion mass is above 600 MeV
and the pion cloud surrounding the nucleon is strongly suppressed. Physical
quantities like the nucleon magnetic moment and axial charge clearly depend strongly
on the pion current, and should therefore be very sensitive to the absence of the full
pion cloud. Furthermore, because of the rapid, nonlinear variation from the chiral logs
arising from Goldstone boson loops, it is clear that a linear extrapolation is completely
inadequate to describe the correct chiral physics.

 In a recent work\cite{Detmold:2001jb}, also discussed by A. Thomas at this
conference, we have shown that chiral extrapolation incorporating the leading
non-analytic behavior from chiral perturbation theory can systematically resolve the
discrepancies in the moments $\langle x^n \rangle_{u-d}$ using the formula:
\begin{equation}
\langle x^n  \rangle_u - \langle x^n  \rangle_d \sim a_n \left[ 1 - {
(3{g_A}^2 +1)m_{\pi}^2 \over (4 \pi f_{\pi})^2} \ln \Bigl( {m_{\pi}^2\over
m_{\pi}^2 + \mu^2} \Bigr) \right] + b_n m_{\pi}^2 \label{chiral_ext}
\end{equation}
The coefficient of the leading non-analytic behavior  $m_{\pi}^2 ln(m_{\pi}^2)$ is
determined from chiral perturbation theory. The parameter $\mu$ specifies the 
scale above which pion loops no longer produce rapid variation. It corresponds to the
upper limit of the momentum integration if one applies a sharp cutoff in the loop
integral and physically corresponds to the inverse size of the quark core of the
nucleon that serves as the source for the pion field. As shown in ref
\cite{Detmold:2001jb} , the value $\mu \sim 550$ MeV, which is consistent with the
value required to extrapolate the nucleon magnetic moment and with chiral nucleon
models, extrapolates the world's supply of lattice data to the experimental values of
$\langle x \rangle_{u-d}$, $\langle x^2 \rangle_{u-d}$, and $\langle x^3
\rangle_{u-d}$. In fig \ref{fit}, we show the extrapolation of our lattice data for
$\langle x \rangle_{u-d}$. From this figure, it is clear that 5\% measurements down to 
$m_{\pi}^2$ = 0.05 GeV$^2$,
would provide data for a definitive
lattice calculation.  
This calculation will require 8 Teraflops-years and thus can be carried out on the next
generation of 10-Teraflops computers.

\section*{ACKNOWLEDGMENTS}

It is a pleasure to acknowledge the contributions of 
D. Dolgov,  S. Capitani, D. Renner, A. Pochinsky, and  R. Brower, to these lattice
calculations,  P. Dreher to the computer cluster infrastructure, 
N. Eicker, T. Lippert,  and  K. Schilling in providing the SESAM configurations,
R. G. Edwards, and U. M. Heller in providing the SCRI configurations, 
W.~Detmold, W.~Melnitchouk,  and A.~W.~Thomas  in
understanding the chiral extrapolation, and M. Stock to the \LaTeX\ manuscript.
\vspace*{-\bigskipamount}


\end{document}